\documentclass[reprint,aps,pre,onecolumn,11pt]{revtex4-2}
\usepackage[utf8]{inputenc}                             
\usepackage[T1]{fontenc}                                
\usepackage{hyperref}                                   
\usepackage{booktabs}                                   
\usepackage{amsfonts,amsmath,amssymb}                   
 \usepackage{bm,graphicx}
\usepackage{color}
\usepackage{mathptmx}
\usepackage{changes,todonotes}
\usepackage{ulem}

\def\bea{\begin{eqnarray}}
\def\eea{\end{eqnarray}}
\def\la{\langle}
\def\ra{\rangle}

\begin{document}

\title{Emergent short-range repulsion for attractively coupled active particles}
\author{Ritwick Sarkar, Urna Basu}
\affiliation{S. N. Bose National Centre for Basic Sciences, Kolkata 700106, India}

\begin{abstract}

We show that heterogeneity in self-propulsion speed can lead to the emergence of a robust effective short-range repulsion among active particles interacting via long-range attractive potentials. Using the example of harmonically coupled active Brownian particles, we analytically derive the stationary distribution of the pairwise distances and reveal that the heterogeneity in propulsion speeds induces a characteristic scale of repulsion between particles. This length scale algebraically increases with the difference in their self-propulsion speeds. In contrast to the conventional view that activity in active matter systems typically leads to effective attraction, our results demonstrate that activity can give rise to an emergent repulsive interaction. This phenomenon is universal, independent of the specific dynamics of the particles or the presence of thermal fluctuations. We also discuss possible experimental realization of this counter-intuitive phenomenon.
\end{abstract}

\maketitle

\section{Introduction}
Active self-propelled particles perform persistent motion by consuming energy at an individual level from their environment~\cite{abp_review,stat_phys_active_matter,active_matter1,RevModPhys.85.1143,abp_review2,time_irre_active_matter,Elgeti_2015}. Examples of such active motion are abundant in nature ranging from bacterial motility~\cite{bergbook} to movement of birds or fish~\cite{giardiana_birdflock,PhysRevLett.75.4326,TONER2005170,PhysRevLett.120.198101,PhysRevE.107.024411}. Artificially designed active agents like Janus particles also exhibit similar motion~\cite{PhysRevE.88.032304,PhysRevLett.123.098001}. The inherently nonequilibrium nature of active particle motion makes their collective properties 
far richer than their equilibrium counterparts. Perhaps the most surprising is the propensity of these particles to form clusters in the absence of any attractive force, leading to a range of unusual phenomena including motility-induced phase separation~\cite{mips1,mips_active_particle,mips_active_rods,vel_alignment_mips}, collective motion~\cite{VICSEK201271,Vicscek_model,PhysRevE.58.4828}, and formation of ordered structure~\cite{active_organization3,PhysRevE.99.032605,PhysRevE.104.024130}.

Theoretical efforts to uncover the origin of the unusual emergent behaviour in active matter often relies on investigating simple model systems comprising few particles. A first step is to explore the effective interaction between two active particles coupled via some simple potential.  The stationary fluctuation of the separation ${\bm r}$ between a pair of passive particles coupled by a sufficiently attractive potential $V(r)$ in two dimensions is governed by the equilibrium radial distribution $P(r) \sim r e^{-V(r)}$. For active particles, however, no such general form exists, and the effective interaction potential, in general, differs from the underlying interaction~\cite{Marconi01092016}.

Recent studies have demonstrated that an effective attractive interaction can emerge between two active particles, even in the presence of hardcore repulsion~\cite{attractive_rtp, twortp_lattice, rtp_dimer_hardcore, gap_statistics_rtp}. A similar attractive interaction emerges also for two particles coupled via long-range attractive potentials~\cite{bound_state_rtp}.  In fact, the presence of persistence may tune an underlying repulsive recoil interaction into an effective attractive one~\cite{attraction_repulsion_active_particle,PhysRevE.107.044134}. 
Furthermore, recent investigations into pairwise gap statistics indicate that similar attractive interactions also emerge for many particle systems~\cite{metson2020jamming}. 
Such emergent attractive interactions are thought to be the driving element behind most of the unusual collective behaviour observed in active particle systems.

In this work, we show that the effective interaction between active particles is not always attractive, as is commonly assumed. In fact, we propose a generic mechanism to generate an effective pairwise repulsive interaction among active particles that are coupled attractively. The crucial element to generate such a repulsion is the presence of heterogeneity in the self-propulsion speed. This remarkable phenomenon is universal and occurs irrespective of the specific active dynamics or presence of thermal fluctuations. Using a simple model, we analytically characterize this repulsive interaction, which, to the best of our knowledge, has not been observed before.

\begin{figure}[t]
     \includegraphics[width=0.95\linewidth]{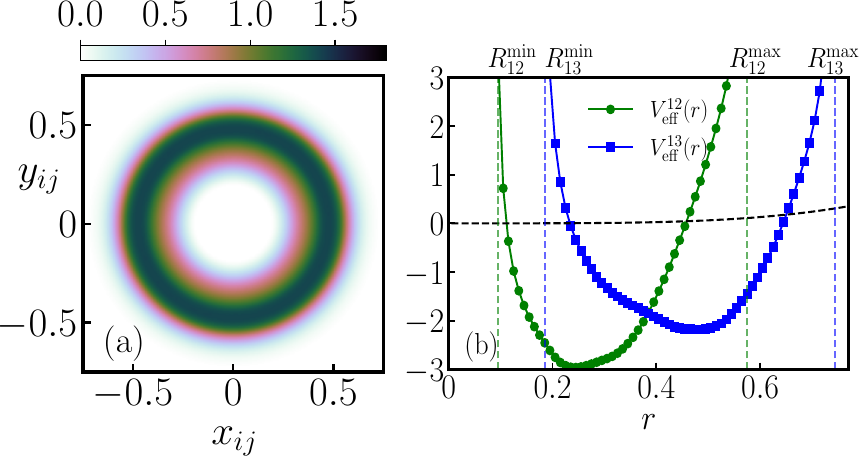}
    \caption{Multi-particle ($N=6$) system interacting via a pair-wise quartic potential $V_q(r)$: (a) Plot of joint probability distribution $P(x_{13},y_{13})$ obtained from numerical simulations. (b) Effective pair-wise potential $V_\text{eff}^{ij}(r)$ extracted from $P(r_{ij})$. Here we have used $v_i=3 i -2$, $\kappa =4$ and $D_i=0.02$.}
    \label{fig:Pxy_q}
\end{figure}

\section{Model and Results}
We consider a collection of $N$ overdamped active Brownian particles (ABP) \cite{abp_2,abp_in_trap,PhysRevE.100.062116} moving in two-dimensions. In the absence of any interaction or external force, each particle self-propels along its internal orientation, which itself evolves stochastically. The set-up comprises particles interacting pairwise via some attractive potential $V(r)$, which depends only on the radial distance $r$ between the two particles. The potential is assumed to be long-ranged and attractive everywhere such that the corresponding force ${\bm f}(r) = - \nabla V(r)$ diverges for $r \to \infty$. The Langevin equations governing the time-evolution of the position ${\bm r}_i = (x_i,y_i)$ of the $i$-th particle is given by,
\bea 
  \dot{\bm{r}}_i(t) &=-\sum_{j\ne i}^{N} {\bm \nabla}_i V(|\bm{r}_i - \bm{r}_j|) + v_i \hat{\bm n}_i(t),\label{eq:lang_eq_1}
\eea 
where, $v_i$ denotes the self-propulsion speed of the $i$-th particle, and $\hat{\bm n}_i = (\cos \theta_i, \sin \theta_i)$ indicates its internal orientation. The orientation of each particle evolves via  a rotational Brownian motion, $\dot \theta_i(t) = \sqrt{2 D_i}\, \eta_i(t)$,
where $\{\eta_i(t) \}$ denote independent white noises with correlation $\la \eta_i(t) \eta_j(t') \ra = \delta_{ij}\delta(t-t')$ and $D_i$ denotes the rotational diffusion coefficient of the $i$-th particle. In the following we consider two specific examples of attractive interactions, namely, harmonic and quartic potentials,
\begin{align}
    V_h(r)=\frac k2 r^2,\quad \mathrm{and}\quad
    V_q(r)=\frac \kappa 4 r^4.\label{form_pot}
\end{align}

We find that, surprisingly, in the strong coupling regime, i.e., when coupling strength is significantly larger than the rotational diffusivity of the particles, an effective short-range repulsion emerges between each pair of particles when their self-propulsion speeds are not equal. This short-range repulsion is manifest in the emergence of a minimum distance between each pair of particles.  An exact expression for the distribution of the pairwise distance $r_{ij} = |\bm{r}_i - \bm{r}_j|$ is derived for the scenario where the particles are coupled to each other via harmonic interaction $V_h(r)$. We show that, in this case, the distance $r_{ij}$ remains bounded in the region,
\begin{align}
R^{\text{min}}_{ij} \equiv \frac {|v_i -v_j|}{Nk} \le r_{ij} \le R^{\text{max}}_{ij} \equiv \frac {(v_i +v_j)}{Nk}, \label{eq:bound_N}
\end{align}
with the distribution satisfying the scaling form,
\bea 
P(r_{ij}) =\frac{1}{R_{ij}^{\text{max}}} {\cal R}_{ij} \left (\frac{r_{ij}}{R_{ij}^{\text{max}}} \right). \label{eq:rij_dist}
\eea
The scaling function is different for each pair of particles,
\bea 
{\cal R}_{ij}(u) = \frac{2 u }{\pi \sqrt{(1- u^2)(u^2 - \ell_{ij}^2)}}, \label{eq:Rij}
\eea 
which depends on the ratio $\ell_{ij}=R_{ij}^{\text{min}}/R_{ij}^{\text{max}}$. 

The corresponding marginal distribution for the $x$-component of the distance $\bm{r}_{ij}$ is also computed exactly, and can be expressed in a scaling form,
\bea 
P(x_{ij}) =\frac{1}{R_{ij}^{\text{max}}} {\cal F}_{ij} \left (\frac{x_{ij}}{R_{ij}^{\text{max}}} \right), \label{eq: marginal} 
\eea 
with the scaling function,
\begin{align}
{\cal F}_{ij}(u) = \begin{cases}
\displaystyle \frac{2i}{\pi^2 \sqrt{\ell_{ij}^2 -u^2}} \Bigg[K\bigg (\frac{1 - u^2}{\ell_{ij}^2 - u^2}\bigg)  \cr 
\displaystyle \, - \sqrt{\frac{\ell_{ij}^2 -u^2}{1 -u^2}} K\bigg (\frac{\ell_{ij}^2 - u^2}{1 - u^2} \bigg) \Bigg] ~\text{for} ~~ |u| < \ell_{ij}, ~~\cr 
\displaystyle \frac{2}{\pi^2 \sqrt{u^2 - \ell_{ij}^2}}  K\Bigg [\frac{1 - u^2}{\ell_{ij}^2 - u^2} \Bigg] ~\text{for} ~~ \ell_{ij} < |u| < 1,
\end{cases} \label{eq:xij_dist}
\end{align}
where $K(u)$ denotes the complete Elliptic integral of the first kind \cite{DLMF}. This marginal distribution is bimodal in shape, with a minimum at $x_{ij} =0 $ and two peaks at $x_{ij} = \pm R_{ij}^{\text{min}}$, which is a signature of the emergent repulsion [see Fig~\ref{fig:diff_v_pr}].  
Equations~\eqref{eq:rij_dist}- \eqref{eq:xij_dist} are one of the central results of this work.

The short-range repulsion also emerges for non-linear coupling. It turns out that, in the strong coupling regime, the distance between two active particles coupled by a generic potential $V(r)$, which is attractive everywhere, must be bounded within the region $R^\text{min}$ and $R^\text{max}$ that satisfies the following relation,
\begin{align}
f(R^\text{min}) =\frac{|v_1 - v_2|}2 ~~ \text{and}~~ f(R^\text{max})=\frac{v_1+v_2}2.\label{eq:cutoff_rad}
\end{align}
Here $f(r)$ denotes the norm of the force acting between the pair of active particles separated by a distance $r$. Similar behaviour is observed for $N>2$ particles coupled with anharmonic potential. This is illustrated in Fig.~\ref{fig:Pxy_q}(a) for a set of particles with pairwise quartic coupling $V_q(r)$. Although analytical results are hard to obtain in this case, the emergent short-range repulsion between the $i$ and $j$-th particles is conveniently characterized via an effective potential~\cite{Marconi01092016},
\begin{align}
    V_\mathrm{eff}^{ij}(r_{ij})=-\log{\left(\frac{P(r_{ij})}{r_{ij}}\right)}+C,
\end{align}
extracted from numerical simulations. Figure~\ref{fig:Pxy_q}(b) shows a plot of this effective potential, which is different for each pair of particles.

In the following we sketch the main computational steps leading to the above results. We start with the case of $N=2$ harmonically coupled active Brownian particles. \\

\subsection{Binary system} To understand the behaviour of the two-particle system, it is convenient to rewrite Eq.~\eqref{eq:lang_eq_1} in terms of the relative distance ${\bm r}={\bm r}_1 - {\bm r}_2$ and center of mass ${\bm q}= \frac 12 ({\bm r}_1+{\bm r}_2)$. For the harmonic coupling $V_h(r)$, we have,
\bea
\dot {\bm r} = - 2k {\bm r} + v_1 \hat {\bm n}_1 - v_2 \hat {\bm n}_2,\quad\text{and}\quad
\dot {\bm q} = v_1 \hat {\bm n}_1 + v_2 \hat{\bm n}_2. \label{eq:r_N2}
\eea 
Clearly, the distance ${\bm r(t)}$ evolves via a Langevin equation Eq.~\eqref{eq:r_N2} which describes the motion of a particle in a harmonic potential of stiffness $2k$, subject to a combination of stochastic noises. While the center of mass undergoes an unbounded motion, the presence of the harmonic trap and bounded nature of the noise ensure that the relative distance ${\bm r}$ would eventually reach a stationary state. In this work, we focus on the behaviour of the stationary state distribution of the distance between the two particles, in the strongly active regime. 

The formal solution of Eq.~\eqref{eq:r_N2} reads,
\bea 
{\bm r}(t) = \int_0^t ds \, e^{-2k (t-s)}\Big[v_1 \hat {\bm n}_1(s) - v_2 \hat {\bm n}_2(s) \Big]. \label{eq:rt_N2}
\eea 
The orientation vectors $\hat{\bm n}_i = (\cos \theta_i, \sin \theta_i)$ evolve independently of ${\bm r}=(x, y)$, eventually reaching a uniform steady state for $\theta_i \in [0, 2\pi]$ for  $t \gg \{D_i^{-1}\}$. On the other hand, the relaxation time-scale of distance ${\bm r}$ in the harmonic trap is given by 
$\tau_k = 1/(2k)$. Thus, in the strongly active regime $k \gg (D_1,D_2)$, the distance vector $\bm r$ relaxes much before the orientations $\hat n_1, \hat n_2$ have changed appreciably. This separation of time-scales implies that, to the first approximation, the complete stationary distribution of ${\bm r}$, in this strongly active regime, can be obtained from the conditional distribution ${\cal P}({\bm r}| \theta_1, \theta_2)$ for fixed $(\theta_1, \theta_2)$, averaging over uniform distributions of $\{\theta_i\}$ in $[0,2\pi]$,
 \bea 
P({\bm r}) = \int_0^{2 \pi} \frac{d\theta_1}{2\pi} \int_0^{2 \pi} \frac{d\theta_2}{2 \pi} {\cal P}({\bm r}| \theta_1, \theta_2). \label{eq:Pr_def1}
\eea 
It is evident from Eq.~\eqref{eq:rt_N2}, for fixed $(\theta_1, \theta_2)$, the distance vector ${\bm r}$ evolves deterministically, leading to,
\bea 
{\cal P}({\bm r}| \theta_1, \theta_2) = \delta\left({\bm r} -  \frac 1{2k}\left[v_1 \hat {\bm n}_1 - v_2 \hat {\bm n}_2 \right] \right). \label{eq:Pr_def2}
\eea 
Consequently, for fixed $(\theta_1, \theta_2)$, the radial distance $r = |{\bm r}|$ between the two particles reaches a constant value 
\begin{align}
r=\frac {|v_1 \hat {\bm n}_1 - v_2 \hat {\bm n}_2|}{2k} = \frac 1{2k}\sqrt{v_1^2 + v_2^2 - 2 v_1 v_2 \cos(\theta_1-\theta_2)}.\label{eq:rad_dist}
\end{align}
Clearly, $r$ is minimum when $\theta_1 - \theta_2 =0$, i.e., orientations of the particles are parallel and $r$ is maximum when $\theta_1 - \theta_2 = \pi$, i.e., the particles are orientated opposite to each other.  Thus, for all possible values of $\theta_1, \theta_2$, the relative distance remains bounded in the regime given by Eq.~\eqref{eq:bound_N} with $N=2$. The above equation implies that particles with different self-propulsion speeds  ($v_1 \ne v_2$)  must be separated by a minimum distance---the strongly active nature of the dynamics gives rise to a short-range repulsive interaction, despite the strong attractive harmonic coupling.

\begin{figure}[t]
		\includegraphics[width=\linewidth]{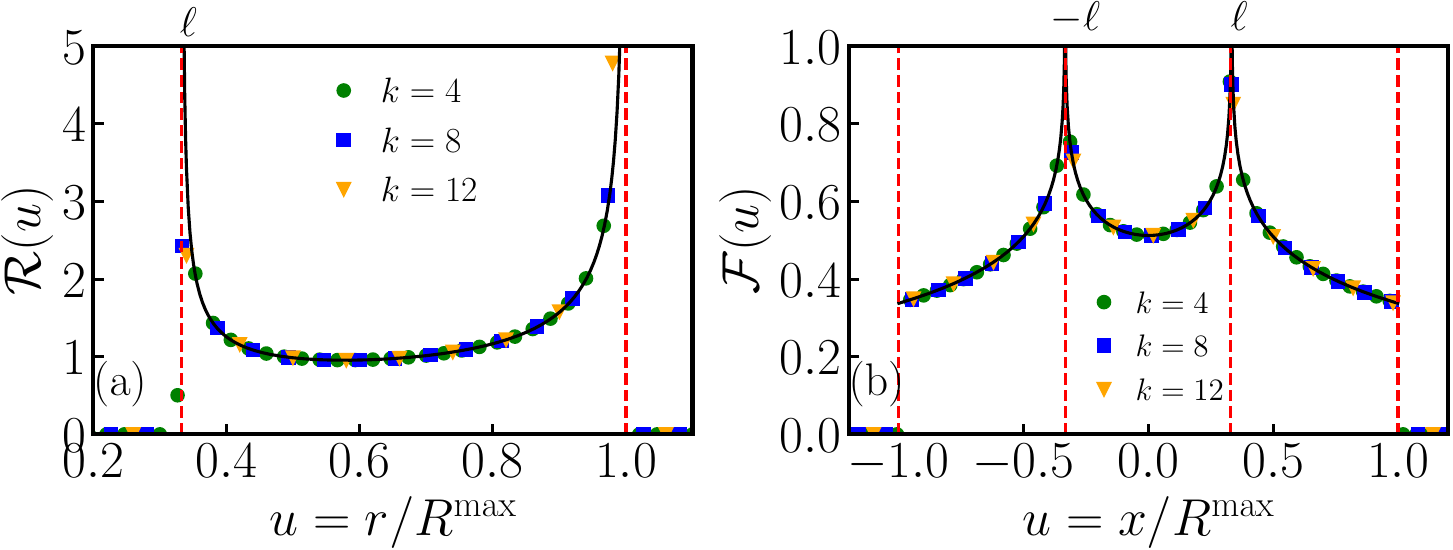}
		\caption{Two harmonically coupled ABP: Probability distribution of (a) the radial distance and (b) the $x$-component of separation ${\bm r}$ obtained from numerical simulations with $v_1=2$, $v_2=4$ and $D_1=D_2=0.01$. The solid black lines in (a) and (b) correspond to the scaling functions \eqref{eq:Rij} and  \eqref{eq:xij_dist}, respectively.}
	\label{fig:diff_v_pr}
\end{figure}

Equations~\eqref{eq:Pr_def1}-\eqref{eq:Pr_def2} allow us to compute the marginal distribution of the radial distance $r = |{\bm r}|$ exactly [see Appendix~\ref{app_dist} for the details]
\bea 
P(r) = \frac{2 r}{\pi \sqrt{[({R^\text{max}})^2 - r^2][r^2 - ({R^\text{min}})^2]}},\label{eq:rdist_N2}
\eea 
for $R^\text{min} < r < R^\text{max}$ which is normalized as $\int P(r) dr =1$. Note that, Eq.~\eqref{eq:rdist_N2} is equivalent to Eqs.~\eqref{eq:rij_dist}-\eqref{eq:Rij} for $N=2$. It should be emphasised that, in the strong-coupling regime, the stationary distributions are independent of the active time-scales $\{D_i^{-1}\}$. Figure~\ref{fig:diff_v_pr}(a) shows the theoretical prediction Eq.~\eqref{eq:rdist_N2} along with the radial distribution obtained from numerical simulations which validates our prediction [see Movie~1(a) in the ESI].

\begin{figure}[t]
    \centering
    \includegraphics[width=\linewidth]{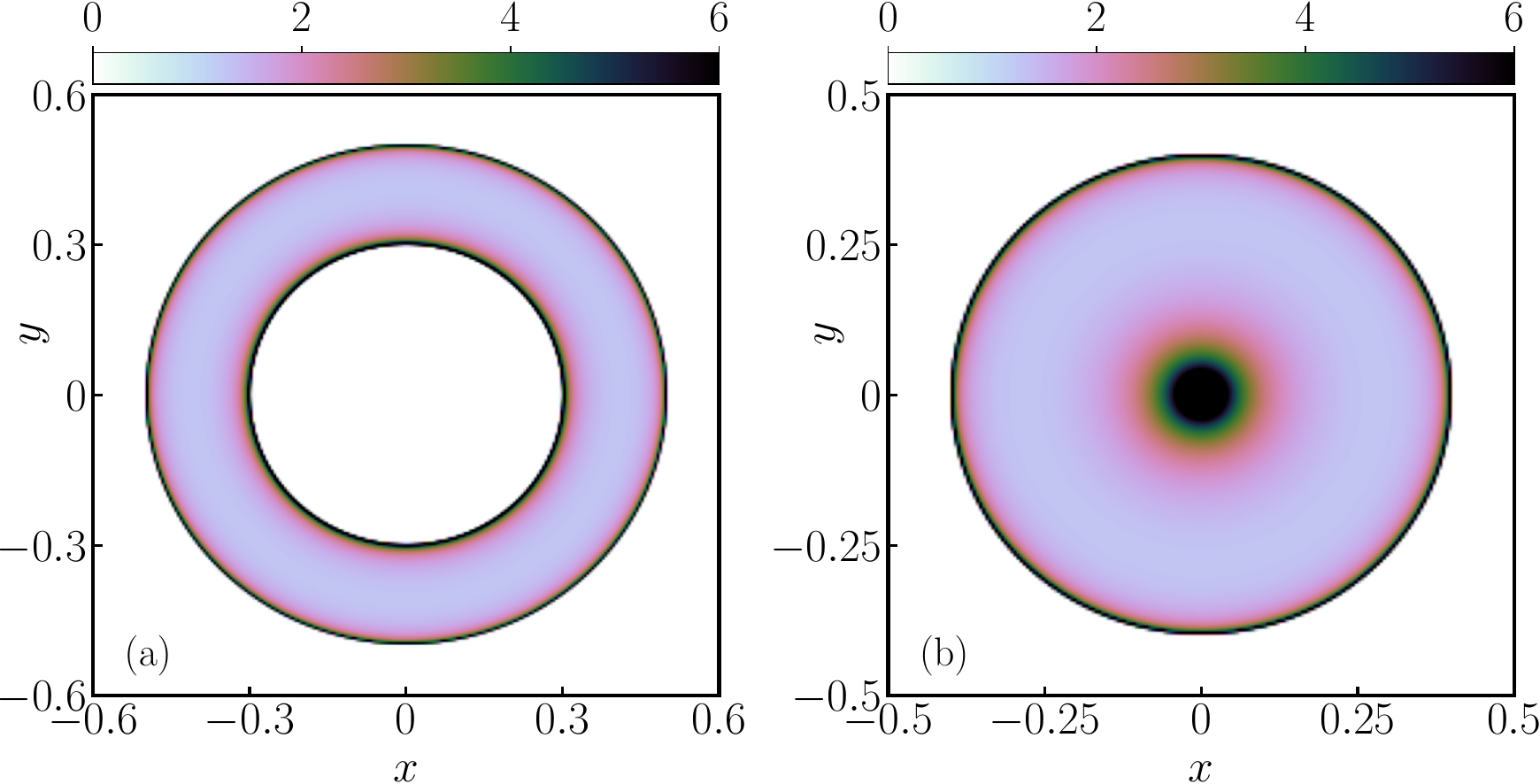}
    \caption{Joint distribution $P(x,y)$ for $N=2$  harmonically coupled particles obtained from numerical simulations for(a) $v_1=1$ and $v_2=4$  and (b) $v_1= v_2=4$. Other parameters are $k=5$ and $D=0.01$. The colour scale represents the value of the probability density $P(x,y)$.}
    \label{fig:pxy_binary}
\end{figure}

The rotational symmetry of the problem allows us to obtain the full two-dimensional distribution $P(x,y)$ which is illustrated in Fig.~\ref{fig:pxy_binary}(a). The empty circular region around the origin ($r < R^\text{min}$) indicates the dynamically inaccessible area of the phase space arising due to the repulsion. Radius of the dynamically inaccessible region $R^\mathrm{min}$, which serves as an alternative measure of range of repulsion, increases linearly with the difference between the self-propulsion speed $v_1-v_2$ [see Eq.~\eqref{eq:bound_N}].

The signature of this emergent repulsion is also visible in $P(x)$, the marginal distribution of the $x-$component of the distance vector ${\bm r}$. This can also be computed analytically and is given by the scaling form [see Appendix~\ref{app_dist} for the details],
\bea 
P(x) =\frac{1}{R^{\text{max}}} {\cal F}_{1\,2} \left (\frac{x}{R^{\text{max}}} \right),\label{eq:xdist_N2}
\eea 
where the scaling function ${\cal F}_{1\,2}(u)$ is given in Eq.~\eqref{eq:xij_dist} with $\ell=R^{\text{min}}/R^{\text{max}}$. This marginal distribution is double peaked, with logarithmic divergences near the peaks at $x = \pm R^\text{min}$, the emergent repulsion giving rise to a minimum at $x=0$. Figure \ref{fig:diff_v_pr}(b) illustrates the behaviour of $P(x)$ along with the same obtained from numerical simulations which show excellent agreement. 

To characterize a nonequilibrium stationary state, it is sometimes useful to define a `nonequilibrium potential', assuming that the stationary state has a Boltzmann form with such an effective potential~\cite{Seifert_2012,attractive_rtp,Marconi01092016,eff_pot_1,Pototsky_2012}. 
In our case, the effective potential can be extracted from the stationary radial separation distribution as, 
\bea
V^{12}_\text{eff}(r) = - \log  \frac {P(r)}{r} +C
\eea 
where $P(r)$ is given by Eq.~\eqref{eq:rdist_N2} and  $C$ is some arbitrary constant. It is evident from Eq.~\eqref{eq:rdist_N2} that  $V_\text{eff}(r) \to \infty$ for $r < R_\text{min}$, consistent with the emergence of short-range repulsion.

To understand the physical origin of this repulsion, let us first note that the bounds of the distribution $R^\text{max}$ and $R^\text{min}$ correspond to the maximum and minimum of 
the turning points of the Langevin equation Eq.~\eqref{eq:r_N2} for 
all possible values of $\theta_1,\theta_2$. For $r > R^\text{max}$, the net force is always negative, pulling the particles closer. In contrast, for $r < R^\text{min}$ the force is positive, pushing the particles apart. This push is generated solely due to the difference in the self-propulsion speeds of the two particles.

In fact, for the special case, $v_1 = v_2$, $R^\text{min}=0$ and Eqs.~\eqref{eq:rdist_N2} - \eqref{eq:xdist_N2} reduce to 
\bea 
P(r) &=& \frac 2{\pi \sqrt{v_0^2/k^2 - r^2}}~~ \text{for} ~~r \le \frac{v_0}k,  \label{eq:equal_vel_rad}\\
P(x) &=& \frac 2{\pi^2|x|}K\left(1 - \frac{v_0^2}{k^2 x^2}\right) ~\text{for}~ |x| < v_0/k.  \label{eq:equal_vel_marginal}
\eea 
 The above equations indicate that the dynamically inaccessible region disappears when both the particles self-propel with the same speed; see Fig.~\ref{fig:pxy_binary}(b) and Fig.~\ref{fig:eq_v_linear} in the Appendix~\ref{app_dist}. In fact, in this case, $P(x)$ shows a logarithmic divergence near $x=0$ indicating an increased probability of the particles being close to each other. In this case, the effective potential $V_\text{eff} (r) \to - \infty$ as $r \to 0$, indicating the enhanced attractive interaction compared to the underlying harmonic coupling. Such enhanced attraction has been observed earlier in case of attractively coupled run-and-tumble particles in one dimension, moving with the same speeds~\cite{bound_state_rtp}.

It is well-known that a single active particle confined in an attractive confining potential is most likely to be away from the minimum of the potential in the strongly active regime~\cite{confinement_janus,confinement_2,trap_range,confinement_janus2,abp_in_trap,abp_trap_1,abp_trap_2,abp_trap_3}.  It should be emphasized here, that the short-range repulsion observed here is physically very different than this phenomenon, and emerges solely due to the heterogeneity in the self-propulsion speeds of the particles. In fact,  two active particles moving with the same speed, coupled via long-range attractive potentials, feel an stronger attraction, as has already been observed~\cite{bound_state_rtp}.

Next, we extend our analysis for more general, non-linear attractive interaction potentials. In this case, the Langevin equation for distance vector becomes,
\bea 
\dot {\bm r} = -2 {\bm f}(r) + v_1 \hat {\bm n}_1 - v_2 \hat {\bm n}_2, \label{eq:rN2_gen}
\eea 
where ${\bm f}(r) = {\bm \nabla} V(r)$. The centre of mass, however, still evolves following the second equation in~\eqref{eq:r_N2}. It is expected that, for sufficiently strong attractive force ${\bm f}(r)$, the distance between the particles will reach a stationary state. 

While it is not possible to solve Eq.~\eqref{eq:rN2_gen} explicitly for arbitrary ${\bm f}(r)$, we can compute the bounds on $r$  from the turning points of the Eq.~\eqref{eq:rN2_gen}, in the strongly active regime,. As discussed before, in the strongly active regime the active time-scale is much larger than the relaxation time-scale in the potential. Hence, one can assume that the orientations do not change appreciably in the time ${\bm r}(t)$ relaxes in the potential. For fixed $\theta_i$, ${\bm r}$ relaxes to the turning point of Eq.~\eqref{eq:rN2_gen},  given by the solution of the equation ${\bm f}(r^*)=  (v_1 \hat n_1 - v_2 \hat n_2)/2$. This equation has a unique solution when $|{\bm f}(r)|$ is a monotonic function of $r$, which is true for the attractive potentials we are considering. Then, the support of the stationary distribution $P(r)$ can be obtained by considering the maximum and minimum of the turning point $r^*$ over all possible orientations $(\theta_1, \theta_2)$, which are given by Eq.~\eqref{eq:cutoff_rad} [see Appendix~\ref{app_sep} for the details].
Hence, for $v_1 \ne v_2$, there is always a minimum distance between the particles, indicating the emergence of the repulsion. Similar to the harmonic case, the point of nearest approach corresponds to the case $\theta_1 - \theta_2 = 0$. The corresponding stationary distributions $P(r)$ and $P(x)$, however, depend on the specific form of the potential. Movie 1(b) in ESI illustrates the stationary state dynamics of two ABPs coupled by $V_q(r)$ [see Eq.~\eqref{form_pot}]. 

It is important to note that, although we have used the example of ABP so far, the results obtained here are universal and hold for other active particle models such as Run-and-Tumble Particles~\cite{rtp_1,rtp_2} or direction reversing ABP~\cite{drabp_1,drabp_2}. This is because, to show the existence of the minimum separation, we have only used the fact that in the strong coupling regime, the orientation evolves much slower than the relaxation of position in the potential. Thus, the details of specific active dynamics do not play any role in this regime.
Moreover, the emergent repulsion also survives in the presence of thermal noise; see Fig.~\ref{fig:dist_temp} in Appendix~\ref{thermal_noise}. \\

\subsection{Multi-particle system} It is intriguing to see whether this emergent repulsion survives for a many particle scenario. First, we consider the harmonic case, where,  $\bm{r}_{ij} \equiv \bm{r}_i - \bm{r}_j$ evolves according to the Langevin equation [see Eq.~\eqref{eq:lang_eq_1}],
\bea 
\dot{\bm{r}}_{ij} = - N k \hat{\bm{r}}_{ij} + v_i \hat{\bm{n}}_i - v_j \hat{\bm{n}}_j. \label{multi_particle}
\eea 
Since the self-propulsion directions $\{ {\bm n}_i \}$ are independent, for each pair, the above equation is equivalent to Eq.~\eqref{eq:r_N2} with a renormalized coupling constant $\tilde k = Nk/2$. Then, it is straightforward to show that, in the strongly active regime $k \gg \{D_i \}$, the relative distance $r_{ij}$ between for each pair of particles $(i,j)$, remains bounded in the regime defined in Eq.~\eqref{eq:bound_N}.
Evidently, the stationary distributions $P(r)$ and $P(x)$  can be obtained from Eqs.~\eqref{eq:rij_dist}-\eqref{eq:Rij} and Eqs.~\eqref{eq: marginal}-\eqref{eq:xij_dist} respectively.
 
It should be emphasized that these distributions hold true for each pair of particles $(i,j)$. Hence, for $N$ particles with distinct self-propulsion speeds, each particle will maintain a minimum distance from all other particles, depending on their speed difference. Seen from the centre of mass of the particles, this amounts to a finite region of the phase space being excluded.

The most interesting case is when $N>2$ particles are coupled via pairwise non-linear force. Then the Langevin equation for ${\bm r}_{ij}$ also depends on ${\bm r}_{im}$ for all possible values of $m$, and it is not possible to obtain the turning points in a straightforward manner, even in the strong-coupling limit. However, physically one still expects a similar behaviour for any long-range attractive coupling [see Movie~1(c) in the ESI]. We illustrate this via numerical simulations. Figure~\ref{fig:Pxy_q}(a) shows a plot of the stationary distribution $P(x_{ij},y_{ij})$ for a fixed pair $(i=1,j=2)$ for $N=6$ particles interacting via the quartic potential $V_q({\bm r})$. The emergent short-range repulsion is evident from the excluded circular region near the origin.

\begin{figure}[h]
    \centering
    \includegraphics[width=\linewidth]{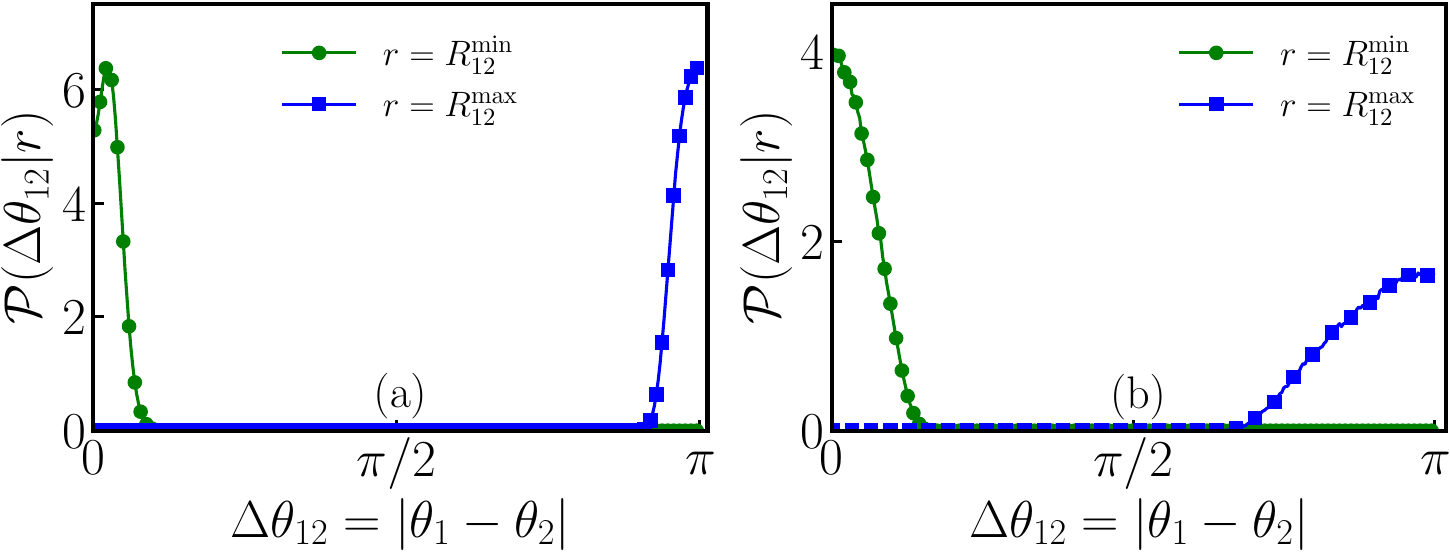}
    \caption{Probability distribution of the relative orientation $\Delta \theta_{ij}$ for minimum and maximum separations between two particles 
        for a system of $N=6$ particles coupled via (a) harmonic potential with $k=2$ and (b) quartic potentials with $\kappa=4$. The self-propulsion speed of the $i$-th particle is $v_i=2i-1$ and the rotational diffusivity of all the particles are fixed at $D_i=0.02$. 
    }
    \label{fig:theta_dist_nonlin_multip}
\end{figure}

The relative orientations for the multi-particle system is also worth investigating. Figure~\ref{fig:theta_dist_nonlin_multip} shows the distribution of the relative orientation $\Delta \theta_{ij}= \theta_i - \theta_j$ at the minimum and maximum separations between a pair of particles, measured from numerical simulations. The peaks near $\Delta \theta_{ij}=0$ for $r_{ij}=R_{ij}^\text{min}$ and $\Delta \theta_{ij}=\pi$ for $r_{ij}=R_{ij}^\text{max}$ confirms that the physical origin of the repulsion lies in the effective two-particle interaction. \\

\section{Conclusion}
Our study provides theoretical evidence of an emergent short-range repulsion among attractively coupled active particles, driven by heterogeneity in their self-propulsion speeds.
Through detailed analytical characterization of the stationary distributions of particle distances, we demonstrate that this repulsion results in a minimum separation between particles, which is directly tied to the difference in their propulsion speeds and is associated with parallel particle orientations. The novel emergent repulsion may be experimentally verified using artificial active particles like self-propelled robots~\cite{hexbug1,hexbug2}, where the control over propulsion speeds is feasible.

Assembly of active particles are typically associated with effective attractive interactions. Presenting a scenario where activity leads to a repulsive interaction, our work paves the way for exploring a broad array of problems in the study of interacting active particles. It is also worthwhile to investigate whether the emergent repulsion is present when the underlying interaction is short-ranged or has a repulsive component, as in Lennard-Jones potential. 
Moreover, it would be interesting to see the effect of speed heterogeneity on collective phenomena like cluster formation and motility induced phase separation. In fact,  numerical evidence provided in a recent work~\cite{D1SM01009C} suggests that diversity in propulsion speed opposes cluster formation. Presence of attractive interaction is also shown to hinder the formation of clusters in active systems~\cite{suppress_phase,mips_self_prop_disk,mips_machn_learn,nonexist_mips,Ray_2024}. It would be worthwhile to see if the simple theoretical model studied here can emulate these phenomena.

\section*{Acknowledgements}
UB acknowledges P. K. Mohanty for useful discussions. RS acknowledges CSIR Grant No. $09/0575(11358)/2021$-EMR-I. UB acknowledges support from the Anusandhan National Research Foundation (ANRF), India, under a MATRICS grant [No. MTR$/2023/000392$].

\appendix
\section*{Appendix}

\section{Stationary separation distribution for harmonic coupling}\label{app_dist}

In this section, we provide the detailed derivation leading to the stationary distribution $P(r)$ and $P(x)$ quoted in Eqs.~\eqref{eq:rij_dist}- \eqref{eq:xij_dist}. We start with the radial distribution $P(r)$ for $N=2$ particle case. In the strong-coupling regime, for fixed $(\theta_1, \theta_2)$, we have from Eq.~\eqref{eq:rad_dist},
\begin{align}
{\cal P}(r| \theta_1, \theta_2) = \delta\left( r -  \frac 1{2k}\sqrt{v_1^2+v_2^2-2 v_1 v_2 \cos{(\theta_1-\theta_2)}} \right).  \label{eq:Pr_def3}
\end{align} 
To obtain $P(r)$ we need to average over all possible values of $\theta_1, \theta_2$ drawn from independent uniform distributions in $[0,2\pi]$ [see Eq.~\eqref{eq:Pr_def1}]. To this end, we first note that for two independent and identically distributed uniform random variables $\theta_1$ and $\theta_2$, the variable $s=\cos (\theta_1 - \theta_2)$ has the probability distribution,
\bea 
\rho(s) = \frac 1{\pi \sqrt{1-s^2}}, \label{eq:rho}
\eea 
\begin{figure}[h]
	\includegraphics[width=\linewidth]{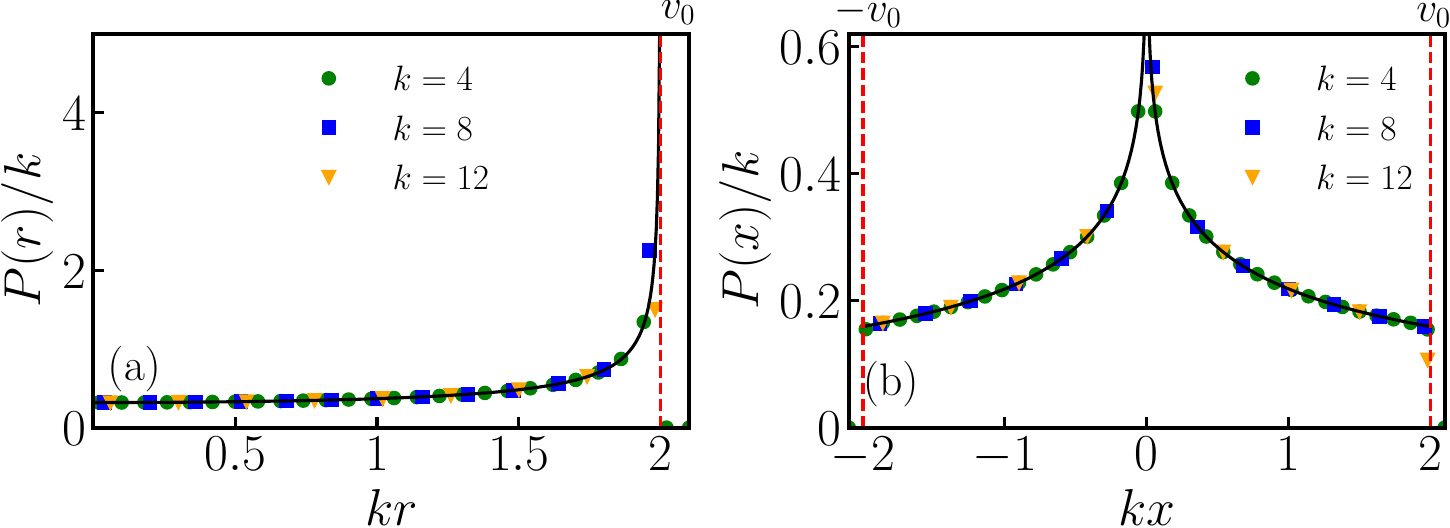}
	\caption{Two harmonically coupled ABP with $v_1=v_2$: Scaled probability distribution of (a) the radial distance and (b) the $x$-component of separation ${\bm r}$ obtained from numerical simulations. The black solid lines in (a) and (b) correspond to Eq.~\eqref{eq:equal_vel_rad} and Eq.~\eqref{eq:equal_vel_marginal}, respectively. Here we have taken $v_1=v_2=2$ and $D_1=D_2=0.01$.}
    \label{fig:eq_v_linear}
\end{figure}
which is supported over the region $-1 \le s \le 1$. Using the above equation in Eq.~\eqref{eq:Pr_def1}, we have,
\bea 
P(r) = \int_{-1}^{1} \frac {ds}{\pi \sqrt{1 - s^2}} \delta\left(r - \frac 1{2k}\sqrt{v_1^2+v_2^2-2 v_1 v_2 s}\right).
\eea 
Evaluating the $s$-integral, we finally get, 
\bea
P(r)=\frac{4 k^2 r}{\pi v_1 v_2 \sqrt{1-{s^{*}}^2} }.\label{rad_dist}
\eea
where, $s^*=(v_1^2+v_2^2-4 k^2r^2)/(2 v_1 v_2) $ denotes the point in the region $[-1,1]$ where the argument of the delta-function vanishes. The explicit form of $P(r)$ is quoted in Eq.~\eqref{eq:rdist_N2}.  For $v_1=v_2=v_0$, $s^* = 2 k^2 r^2/v_0^2$ and the radial distribution reduces to  Eq.~\eqref{eq:equal_vel_rad}.

Next, we compute the marginal distribution of the $x$-component of the separation vector ${\bm r}$. In the strong-coupling limit, for fixed $(\theta_1, \theta_2)$, we have from Eq~\eqref{eq:rt_N2},
\bea
x &=& a_1 \cos \theta_1 - a_2 \cos \theta_2, 
~\text{where,}~ a_i=\frac{v_i}{2 k}. \label{x_y_dynamics}
\eea 
The conditional distribution ${\cal P}(x| \theta_1, \theta_2)$  is then given by,
\bea
{\cal P}(x| \theta_1, \theta_2)=\delta(x - a_1 \cos \theta_1 + a_2 \cos \theta_2).
\eea 

The marginal probability distribution of $x$-component, thus, is given by,
\begin{align}
P(x) = \int_{-\infty}^{\infty} dz_1 \int_{-\infty}^{\infty} dz_2 \,  \delta(x - a_1 z_1 + a_2 z_2) \rho(z_1) \rho(z_2),~~\label{eq:Px_int}
\end{align} 
where $z_i=\cos{\theta_i}$. It is straightforward to show that the distribution of $z_i$ is nothing but $\rho(z_i)$, given in Eq.~\eqref{eq:rho}.  Using Eq.~\eqref{eq:rho} in Eq.~\eqref{eq:Px_int} and integrating over $z_2$, we arrive at,
\begin{align}
P(x) = \frac 1{a_1 \pi^2} \int_{-1}^{1} d z_1  \frac{\Theta(z_1 - b_1)}{\sqrt{1-z_1^2}} \frac{\Theta(b_2 -z_1)}{\sqrt{(z_1-b_1)(b_2-z_1)}},
\label{marginal_integral}
\end{align}

where $\Theta(z)$ is the Heaviside theta function. We have also defined $b_1=(x-a_2)/a_1$ and $b_2=(x+a_2)/a_1$ for notational simplicity. The above integral can be evaluated exactly and leads to, 
\begin{align}
P(x) = \begin{cases}
\displaystyle \frac{2i}{\pi^2 \sqrt{(R^\text{min})^2 -x^2}} \Bigg[K\bigg (\frac{(R^\text{max})^2 - x^2}{(R^\text{min})^2 - x^2}\bigg)  \cr 
\displaystyle \, - \sqrt{\frac{(R^\text{min})^2 -x^2}{(R^\text{max})^2 -x^2}} K\bigg (\frac{(R^\text{min})^2 - x^2}{(R^\text{max})^2 - x^2} \bigg) \Bigg]\cr
\quad\quad\quad\quad\quad\quad\quad\quad\quad\quad~\text{for} ~~ |x| < R^\text{min}, ~~\cr 
\displaystyle \frac{2}{\pi^2 \sqrt{x^2 - (R^\text{min})^2}}  K\Bigg [\frac{(R^\text{max})^2 - x^2}{(R^\text{min})^2 - x^2} \Bigg]\cr
\quad \quad\quad\quad\quad\quad\quad\text{for} ~~ R^\text{min} < |x| < R^\text{max},
\end{cases}
\end{align}
where $K(u)$ denotes the complete Elliptic integral of the first kind \cite{DLMF}. This distribution shows a logarithmic divergence near the peaks at $x= \pm R^\text{min}$,
\begin{align}
    P(x)\simeq \frac{-\log{|x-R^\text{min}|}+\log{\left[(R^\text{max})^2-(R^\text{min})^2\right]}}{\sqrt{(R^\text{max})^2-(R^\text{min})^2}}.
\end{align}

\begin{figure}[t]
    \centering
    \includegraphics[width=\linewidth]{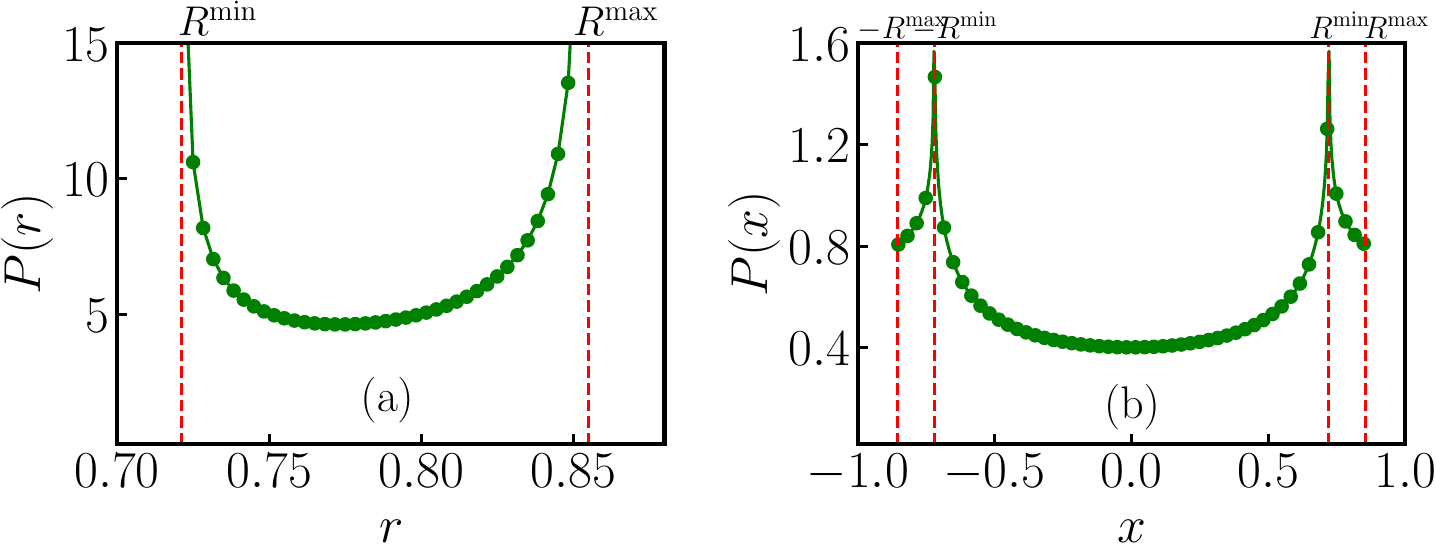}
    \caption{Two ABP coupled with quartic potential $V_q({\bm r})$: Probability distribution of (a) the radial distance and (b) the $x$-component of separation ${\bm r}$ obtained from numerical simulations with the strength of the potential $\kappa=4$, $v_1=1,v_2=4$ and $D_1=D_2=0.02$. The red dashed line corresponds to the range of the distribution given by Eq.~\eqref{quad_dimer_range}.}
    \label{fig:quad_dist_dimer}
\end{figure}
For the special case $v_1 = v_2=v_0$, we have  $R^{\text{min}}=0, R^{\text{max}} = v_0/k$. The marginal distribution, in this case, is given by,
\begin{align}
P(x)= 
\begin{cases}
\displaystyle \frac{2k}{\pi^2 v_0} \int_{2 k x/v_0-1}^{1} \frac {dz}{\sqrt{1-z^2}\sqrt{(z-c_1)(c_2-z)}}\cr \quad \quad \quad \quad\quad \quad\quad\quad\quad\quad\quad\quad\quad\quad \text{for } 0<x\leq \frac{v_0}{k}, ~~\cr
\displaystyle \frac{2k}{\pi^2 v_0} \int_{-1}^{2 k x /v_0+1} \frac {dz}{\sqrt{1-z^2}\sqrt{(z-c_1)(c_2-z)}}\cr\quad \quad \quad \quad\quad \quad\quad\quad\quad\quad\quad\quad\quad\quad  \text{for } -\frac{v_0}{k}\leq x<0,
\end{cases}
\end{align}
where, $c_1=(2 k x-v_0)/v_0$ and $c_2=(2 kx+v_0)/v_0$. After performing the integral we arrive at Eq.~\eqref{eq:equal_vel_marginal}. The distribution in  Eq.~\eqref{eq:equal_vel_marginal} shows a logarithmic divergence near $x=0$, 
\bea 
P(x) \simeq \frac{2k}{\pi^2 v_0} \Big(- \log x + \log \frac{4v_0}k \Big).
\eea 

The radial and $x$-marginal distributions for the $v_1=v_2$ case are illustrated in Fig.~\ref{fig:eq_v_linear}.

\section{Anharmonic coupling}\label{app_sep}

\begin{figure}[h]
    \centering
    \includegraphics[width=0.95\linewidth]{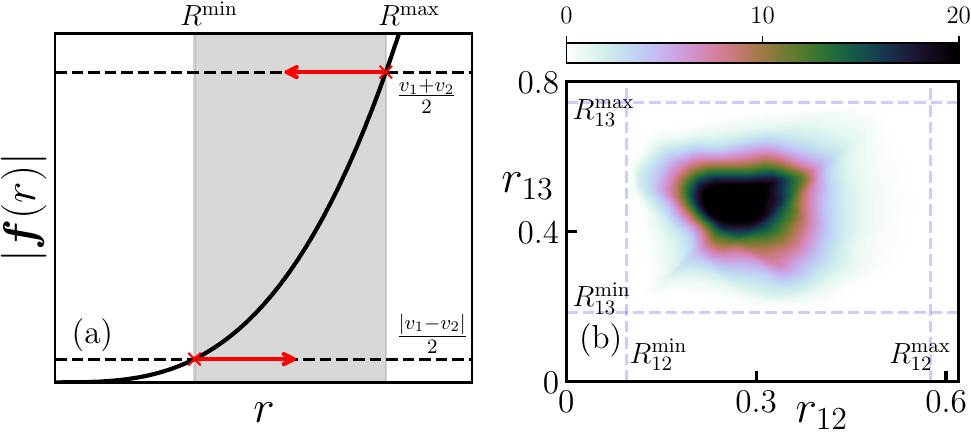}
    \caption{(a) Schematic of the turning points: The black solid line represents the attractive force $|{\bm f}(r)|$, dashed lines mark the bounds of the active force. The red crosses indicate the turning points and the grey-shaded region indicates the accessible region. The arrow arrows show the direction of the force at the turning points. (b) Joint probability distribution $P(r_{12}, r_{13})$ for $N=6$ particles interacting with quartic potential $V_q(r)$. The lower and upper bounds of $r_{12}$ and $r_{13}$ shown in the plots are extracted from numerical simulations. The parameters used in the simulation are $v_i=3 i -2$, $\kappa =4$ and $D=0.02$. }
    \label{fig:phase_schm_turningpoint}
\end{figure}
The equation of motion of the radial distance between two ABPs coupled to each other by a long-range attractive force $\bm{f}(r)$ is given by 
\bea 
\dot {\bm r}_{12} = -2 {\bm f}(r_{12}) + v_1 \hat {\bm n}_1 - v_2 \hat {\bm n}_2, 
\eea 
where $\hat{\bm n}_i = (\cos \theta_i, \sin \theta_i)$ indicates the internal orientation of the ABPs. In the strong coupling limit, the distance vector $\bm r$ relaxes in the attractive potential much before the orientations $\hat n_1, \hat n_2$ have changed appreciably. For any fixed $(\theta_1, \theta_2)$, the radial distance eventually approaches the turning point  ${ r}^*$ [See Fig.~\ref{fig:phase_schm_turningpoint}(a)], which satisfies,
\begin{align}
|{\bm f}( r^*)|=\frac 12 \sqrt{[v_1^2+v_2^2-2 v_1 v_2 \cos{(\theta_1-\theta_2)}]}.\label{eq:theta_d}
\end{align}
Clearly, for monotonically increasing functions $|{\bm f}(r)|$, which we are considering, Eq.~\eqref{eq:theta_d} implies Eq.~\eqref{eq:cutoff_rad} quoted in the main text. Especially, for the quartic potential  $V_q(r)$ [see Eq.~\eqref{form_pot}], we have,
\begin{align}
R^\text{min} = \left(\frac{|v_1-v_2|}{2 \kappa}\right)^{\frac{1}{3}},\quad\text{and}\quad R^\text{max}=\left(\frac{v_1+v_2}{2 \kappa}\right)^{\frac{1}{3}}.\label{quad_dimer_range}
\end{align}
The probability distribution of the the radial separation ${\bm r}$ and its $x$-component for a pair of ABP coupled with quartic potential is shown in Fig.~\ref{fig:quad_dist_dimer} along with the $R^\text{min}$ and $R^\text{max}$ [see Movie 1(b) in the ESI].

For multi-particle system the signature of the emergent repulsion is also apparent in the joint distribution $P(r_{ij}, r_{im})$. This is illustrated in Figure~\ref{fig:phase_schm_turningpoint}(b) for quartic interaction, which shows that a large area around the origin remains dynamically inaccessible.

\section{Effect of thermal noise}\label{thermal_noise}

In the presence of thermal noise the Langevin equation \eqref{eq:lang_eq_1}
reads,
\begin{align}
  \dot{\bm{r}}_i(t) &=-\sum_{j\ne i}^{N} {\bm \nabla}_i V(|\bm{r}_i - \bm{r}_j|) + v_i \hat{\bm n}_i(t) + \sqrt{2T}~{\bm \zeta}_i(t),~ \label{eq:Lang_eq_T}
\end{align} 
where  $T$ denotes the temperature and $\{ {\bm \zeta}_i(t) =(\zeta^{x}_i(t), \zeta^{y}_i(t)) \}$ are independent white noises with $\la \zeta_i^{\alpha}(t) \zeta_j^{\beta}(t') \ra = \delta_{\alpha\beta}\delta_{ij} \delta(t-t')$. In this scenario, the distance ${\bm r}= {\bm r}_1 - {\bm r}_2$ between $N=2$ harmonically coupled particles evolve according to,
\begin{align}
\dot {\bm r} = - 2k {\bm r} + v_1 \hat {\bm n}_1 - v_2 \hat {\bm n}_2+\sqrt{2T} {\bm \zeta}(t),
\end{align} 
\begin{figure}[t]
    \centering
    \includegraphics[width=\linewidth]{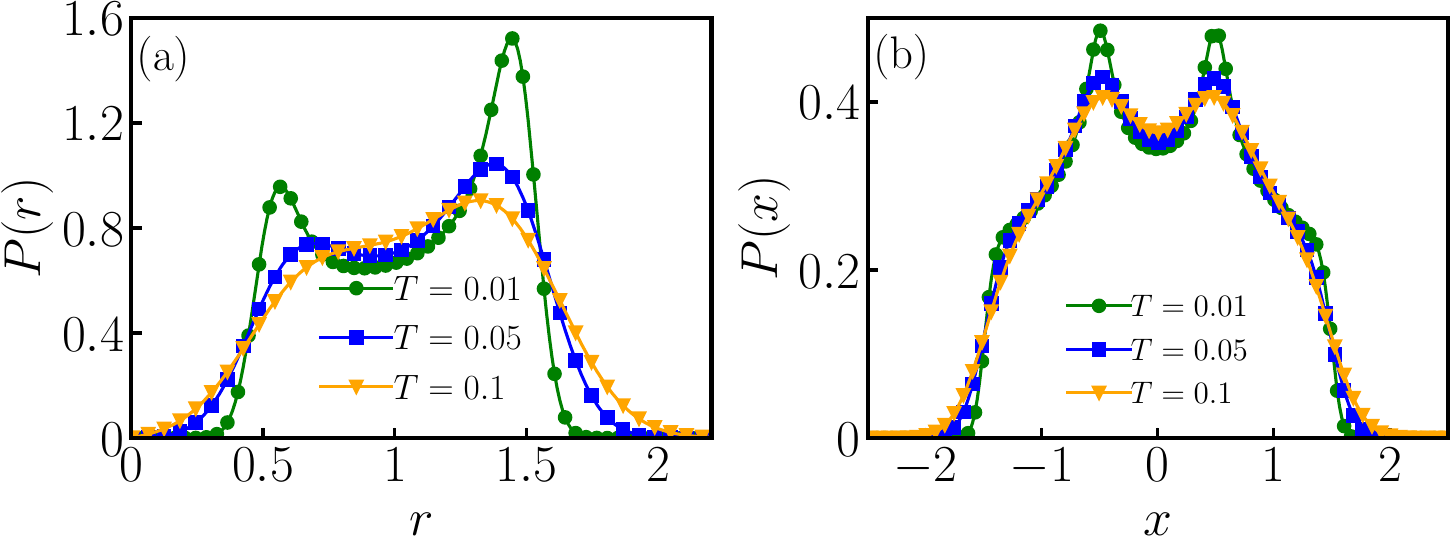}
    \caption{Two harmonically coupled ABP in presence of thermal noise of strength $T$: Probability distribution of (a) the radial distance and (b) the $x$-component of separation ${\bm r}$ obtained from numerical simulations with $k=2$, $D_1=D_2=0.02$, $v_1=2$ and $v_2=4$.}
    \label{fig:dist_temp}
\end{figure}
where ${\bm \zeta}(t) = {\bm \zeta}_1(t) - {\bm \zeta}_2(t)$ is another white noise with  $\la \zeta^\alpha(t) \zeta^\beta(t') \ra = 2 \delta_{\alpha \beta}\delta(t-t')$. The emergent repulsion is expected to survive for small temperatures $T \ll (v_1 - v_2)^2 / k$. Fig.~\ref{fig:dist_temp} shows plots of $P(r)$ and $P(x)$ for different temperatures $T$---the presence of thermal noise leads to a non-zero probability of the particles being separated by distances than $r<R^\text{min}$, however, $P(r)$ and $P(x)$ having peaks at $r=R^\text{min}$ confirms the emergence of repulsion.

\section{Details of numerical simulation}

To numerically simulate the interacting active particle dynamics described by the Langevin equation~\eqref{eq:lang_eq_1}, we discretize it in time-steps of duration $\Delta t$, to first order in $\Delta t$. Thus, the position components $(x_i(t), y_i(t))$ and orientation $\theta_i(t)$ of the $i$-th particle are updated as, 
\begin{align}
    x_i(t+\Delta t) &=x_i(t)-\sum_{i\neq j}^N \frac{x_i(t)-x_j(t)}{|{\bm r}_i(t)-{\bm r}_j(t)|}g(|{\bm r}_i(t)-{\bm r}_j(t)|)\Delta t\cr
    &+v_i \cos\theta_i(t)\Delta t,\label{sim_1}\\
    y_i(t+\Delta t)&=y_i(t)-\sum_{i\neq j}^N\frac{y_i(t)-y_j(t)}{|{\bm r}_i(t)-{\bm r}_j(t)|}g(|{\bm r}_i(t)-{\bm r}_j(t)|)\Delta t\cr
    &+v_i \sin\left[\theta_i(t)\right]\Delta t,\label{sim_2}\\
    \theta_i(t+\Delta t)&=\theta_i(t)+\sqrt{2 D_i \Delta t}\,\eta_t,
\end{align}
where $g(r) = V'(r)$ and $\eta_t$ is a random number drawn from the standard Normal distribution with zero mean and unit variance $\mathcal{N}(0,1)$. For the harmonic coupling $g(r)=k r$ whereas for the 
quartic case, $g(r)=\kappa r^3$.

For all the data presented here, we have used $\Delta t=10^{-3}$ and the distributions are obtained by averaging over at least $10^{9}$ samples.

\bibliographystyle{apsrev4-2}
\bibliography{ref}
\end{document}